\documentclass[aps,pre,preprint,superscriptaddress,showpacs,longbibliography]{revtex4-1}

\usepackage{amsmath}
\usepackage{graphicx}
\usepackage{natbib}
\usepackage{tabularx}
\usepackage{multirow}
\usepackage{slashbox}

\begin{document}


\title{Induced charge electrophoresis of a conducting cylinder in a non-conducting cylindrical pore and its micromotoring application}



\author{Huicheng Feng}
\affiliation{School of Mechanical and Aerospace Engineering, Nanyang Technological University, 50 Nanyang Avenue, Singapore 639798, Singapore}
\author{Teck Neng Wong}
\email{mtnwong@ntu.edu.sg}
\affiliation{School of Mechanical and Aerospace Engineering, Nanyang Technological University, 50 Nanyang Avenue, Singapore 639798, Singapore}
\author{Zhizhao Che}
\affiliation{State Key Laboratory of Engines, Tianjin University, Tianjin 300072, China}



\date{\today}

\begin{abstract}
Induced charge electrophoresis of a conducting cylinder suspended in a non-conducting cylindrical pore is theoretically analyzed, and a micromotor is proposed utilizing the cylinder rotation. The cylinder velocities are analytically obtained in the Dirichlet and the Neumann boundary conditions of the electric field on the cylindrical pore. The results show that the cylinder not only translates but also rotates when it is eccentric with respect to the cylindrical pore. The influences of a number of parameters on the cylinder velocities are characterized in detail. The cylinder trajectories show that the cylinder approaches and becomes stationary at certain positions within the cylindrical pore. The proposed micromotor is capable of working under a heavy load with a high rotational velocity when the eccentricity is large and the applied electric field is strong.
\end{abstract}


\maketitle

\section{Introduction}
Particle and fluid manipulations utilizing electric fields have been extensively studied due to their advantages in various applications, including colloid science \cite{anderson1989colloid}, micro/nanofluidics \cite{bazant2010induced,Zhong2015a}, chemistry \cite{Vilkner2004,West2008}, biology \cite{jiang2010separating,Squires2013}, biomedicine \cite{keren2003protein,Huang2010}, etc. Among these manipulation methods, induced charge electrophoresis (ICEP) is receiving increasingly significant interests because of its great potential in various applications, ranging from the manipulations of droplets \cite{schnitzer2013nonlinear,Schnitzer2013drop} and particles \cite{boymelgreen2012alternating,Feng2015}, to the device development in lab-on-a-chip systems, e.g., micromixers \cite{Daghighi2013,Feng2016pofchaotic}, microvalves \cite{sugioka2010high,daghighi2011micro} and micromotors \cite{squires2006breaking,Boymelgreen2014}. When a conducting (ideally polarizable) particle is subjected to an external electric field, it polarizes immediately. The polarization surface charges attract counterions in the electrolyte solution, establishing an induced electric double layer (EDL). The interactions between the applied electric field and the induced EDL lead to fluid flow, known as induced charge electroosmosis (ICEO). The particle motions due to ICEO is termed as induced charge electrophoresis (ICEP). The ICEP velocity is a quadratic function of the strength of the applied electric field because the zeta potential of the conducting (ideally polarizable) particle is induced by the applied electric field \cite{Yariv2005},
\begin{equation}
\zeta_i=-\phi+\int_A{\phi dA}/A,
  \label{eq1zetaidefinition}
\end{equation}
where $\zeta_i$ is the induced zeta potential of the conducting particle, $\phi$ is the applied electrical potential on the particle surface, and $A$ is the area of the particle surface.

Pioneering studies of the ICEP were carried out in colloid science decades ago \cite{Gamayunov1986,Dukhin1986}. Thanks to the rapid advancement in material science and nanotechnology, which provides various kinds of conducting particles for micro/nanofluidics \cite{Garcia-Sanchez2012,Zhong2014,Zhong2015}, ICEP regains researchers' attention in recent years \cite{bazant2010induced,ramos2011electrokinetics,Feng2016}. As the particles are often bounded or contained in channels or chambers in reality, the boundary effect is of significant importance in the relevant applications. Some studies have been conducted considering the planer wall effect in the ICEP motion of particles \cite{gangwal2008induced,Yariv2009remote,Hamed2009near,kilic2011induced,Sugioka2011}. The repulsion or attraction effects of the planar wall on different particles, ranging from cylinders \cite{zhao2007effect,kilic2011induced}, Janus particles \cite{gangwal2008induced}, spheres \cite{Yariv2009remote,Hamed2009near,kilic2011induced}, to ellipsoids \cite{Sugioka2011}, have been reported. However, in such studies, the walls are all straight and uncharged. Investigations on ICEP behavior of particles near a curved and/or charged non-conducting wall remain limited. In an effort to improve the physical insights of this problem, we hereby carry out a comprehensive study on the ICEP motion of a conducting cylinder suspended in a non-conducting cylindrical pore. The analytical evaluations on the cylinder velocities reveal that the cylinder is not only driven into translation but also, surprisingly, into rotation when the cylinder and the cylindrical pore are eccentric.

The ICEP rotation of the cylinder presents a promising potential as a micromotor, which has long been crucial in the development of micromachines for biomedicine \cite{zhao2013influence}, biochemistry \cite{guix2014nano}, environmental science \cite{gao2014environmental}, etc., thus, remains a hot topic of scientific and technological interests. Various micromotors have been proposed and studied \cite{liu2011autonomous,wu2013self,mou2013self}. Most of such studies are centered on Janus particles \cite{squires2006breaking,mou2013self}, segmented rods with different metals \cite{liu2011autonomous}, and microtubes with layer-by-layer deposited metals \cite{wu2013self}. ICEP rotations of nonspherical particles \cite{Yariv2005,saintillan2006hydrodynamic,yariv2008slender}, Janus \cite{squires2006breaking} particles and cylinders in pair interactions \cite{Feng2016} have been reported. An ICEP micromotor composed of three Janus particles has been proposed and theoretically analyzed years ago \cite{squires2006breaking}. Lately, the ICEP rotation of a doublet Janus particle has been experimentally captured \cite{Boymelgreen2014}. However, all these proposed structures are composed of different materials, which are complicated and bring fabrication challenges. We hereby propose a micromotor utilizing the ICEP rotation of the cylinder in the cylindrical pore, which has advantages of simple geometry and material property. Thus, it is easy to fabricate. The analysis shows that the micromotor is capable of providing a large rotational velocity and bearing a heavy load. The study could contribute to the understanding of the ICEP behavior of a conducting cylinder in non-conducting cylindrical pore, and provide helpful insights in micromotor development.

\section{Mathematical formulation}
\begin{figure}[!htb]
\centering
\includegraphics[width=3.25in]{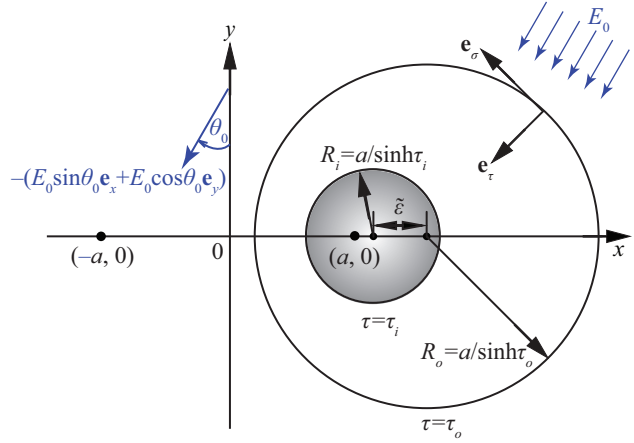}
\caption{Schematic illustration of the conducting cylinder suspended in the cylindrical pore. $(\pm a,0)$ are the two foci of the bipolar coordinates; $\mathbf{e}_{\tau}$ and $\mathbf{e}_{\sigma}$ are the unit vectors in the bipolar coordinates that normal and tangent to the cylinder surface, respectively; $\tau=\tau_i$ and $\tau_o$ indicate the surfaces of the cylinder and the cylindrical pore, respectively; $(a\coth\tau_i,\,0)$ and $(a\coth\tau_o,\,0)$ are the centers of the cylinder and the cylindrical pore, respectively, in the Cartesian coordinates; $R_i=a/{\sinh\tau_i}$ and $R_o=a/{\sinh\tau_o}$ are the radii of the cylinder and the cylindrical pore, respectively; and $\tilde\varepsilon$ is the distance between the centers of the cylinder and the cylindrical pore. The uniform electric field $-(E_0 \sin\theta_0 \mathbf{e}_x+E_0 \cos\theta_0 \mathbf{e}_y)$ is imposed, where the electric field phase angle $\theta_0$ defines the direction of the electric field.}
 \label{fig1}
\end{figure}

A two-dimensional (2D) conducting cylinder is suspended in a non-conducting cylindrical pore filled with an electrolyte solution. A Cartesian coordinate system is introduced in which the centers of the cylinder and the cylindrical pore are on the positive $x-$axis (Fig.\ \ref{fig1}). As the cylinder and the cylindrical pore are commonly eccentric, a bipolar coordinate system is defined in the Cartesian coordinates \cite{keh1991boundary},
\begin{equation}
  x=a\frac{\sinh\tau}{\cosh\tau-\cos\sigma},\quad y=a\frac{\sin\sigma}{\cosh\tau-\cos\sigma},
  \label{eq1bipolarcoord}
\end{equation}
so as to conveniently describe the eccentric geometry (Fig.\ \ref{fig1}). Here $-\infty<\tau<\infty$; $0<\sigma\le 2\pi$; $(\tau, \, \sigma)$ denote the coordinates of the bipolar coordinate system; and $a$ is a positive constant in the bipolar coordinates. The surfaces of the cylinder and the cylindrical pore are indicated by $\tau =\tau_i$ and $\tau_o$, respectively, in the bipolar coordinates.

To quantitatively describe the eccentric geometry, two parameters, i.e., the radius ratio $R_r$ and the eccentricity $\varepsilon$, are introduced,
\begin{equation}
R_r=\frac{R_i}{R_o},\quad \varepsilon=\frac{\tilde{\varepsilon}}{R_o-R_i},
\label{Rr}
\end{equation}
where $R_i$ and $R_o$ are the radii of the cylinder and the cylindrical pore, respectively; $\tilde{\varepsilon}$ is the distance between the centers of the cylinder and the cylindrical pore (Fig.\ \ref{fig1}). When the eccentricity $\varepsilon$ decreases to zero, the cylinder and the cylindrical pore become concentric.

The bulk fluid outside the EDLs is electrically neutral. Thus, the Laplace equation is applied,
\begin{equation}
\nabla^2\phi=0,
  \label{eq2LaplaceEq}
\end{equation}
where $\phi$ is the electrical potential of the bulk fluid.

The electric field lines are expelled by the EDL on the cylinder. Hence, the no-flux condition is applied,
\begin{equation}
\mathbf{e}_{\tau}\cdot\nabla\phi=0 \quad\text{at}\quad \tau=\tau_i,
  \label{eq3noflux}
\end{equation}
where $\mathbf{e}_{\tau}$ is the unit vector normal to the cylinder surface in the bipolar coordinates (Fig.\ \ref{fig1}).

Given the uniformly applied electric field $-(E_0 \sin\theta_0 \mathbf{e}_x+E_0 \cos\theta_0 \mathbf{e}_y)$, the boundary condition on the cylindrical pore is,
\begin{equation}
\phi =E_0 a \frac{\sin\sigma\cos\theta_0+\sinh\tau_o\sin\theta_0}{\cosh\tau_o-\cos\sigma} \quad \text{at} \quad \tau =\tau_o,
  \label{eq4potentialDirichlet}
\end{equation}
using the Dirichlet condition, or
\begin{eqnarray}
\frac{\partial \phi}{\partial \tau} &=&E_0 a \frac{(1-\cosh\tau_o\cos\sigma)\sin\theta_0-\sinh\tau_o\sin\sigma\cos\theta_0}{(\cosh\tau_0-\cos\sigma)^2}
  \nonumber\\
&&  \text{at} \quad \tau =\tau_o,
  \label{eq4potentialNeumann}
\end{eqnarray}
using the Neumann condition. Both conditions lead to the uniform electric field $-(E_0 \sin\theta_0 \mathbf{e}_x+E_0 \cos\theta_0 \mathbf{e}_y)$ in the fluid flow when the cylinder disappears, although Eq.\ (\ref{eq4potentialNeumann}) does not define the tangential electric field on the cylindrical pore. This paper presents the derivation using the Dirichlet condition (Eq.\ (\ref{eq4potentialDirichlet})). For the derivation using the Neumann condition (Eq.\ (\ref{eq4potentialNeumann})), pleases refer to Section C of the Supplementary.

Solving Eq.\ (\ref{eq2LaplaceEq}) together with Eqs.\ (\ref{eq3noflux}) and (\ref{eq4potentialDirichlet}), the electrical potential is obtained,
\begin{eqnarray}
\phi&=&E_0 a \sin\theta_0+ E_0 a \sum_{n=1}^{\infty} \frac{2\text{e}^{-n\tau_o}\cosh n(\tau_i-\tau)}{\cosh n(\tau_i-\tau_o)}
  \nonumber\\
&&  \times(\sin\theta_0\cos n\sigma+\cos\theta_0\sin n\sigma).
  \label{eq5phiDirichlet}
\end{eqnarray}

The zeta potential of the non-conducting cylindrical pore is fixed, $\zeta_f$; while that of the conducting cylinder is induced by the imposed electric field and obtained by substituting Eq.\ (\ref{eq5phiDirichlet}) into Eq.\ (\ref{eq1zetaidefinition}),
\begin{eqnarray}
\zeta_i&=&-E_0 a\sum_{n=1}^{\infty}\frac{2\text{e}^{-n\tau_o}}{\cosh n(\tau_i-\tau_o)}
  \nonumber\\
&&  \times(\sin\theta_0\cos n\sigma+\cos\theta_0\sin n\sigma) \quad \text{at} \quad \tau=\tau_i.
  \label{eq7zetaiDirichlet}
\end{eqnarray}

The tangential electric fields $E_\sigma$ on the cylinder and the cylindrical pore are obtained from Eq.\ (\ref{eq5phiDirichlet}) through $\mathbf{E}=-\nabla\phi$,
\begin{eqnarray}
E_\sigma&=&E_0 (\cosh\tau_i-\cos\sigma)\sum_{n=1}^{\infty}\frac{2n\text{e}^{-n\tau_o}}{\cosh n(\tau_i-\tau_o)}
  \nonumber\\
&&  \times(\sin\theta_0 \sin n\sigma-\cos\theta_0 \cos n\sigma) \quad \text{at} \quad \tau=\tau_i,\quad
  \label{eq7E11innerDirichlet}
\end{eqnarray}
\begin{eqnarray}
E_\sigma&=&E_0 (\cosh\tau_o-\cos\sigma)\sum_{n=1}^{\infty}2n\text{e}^{-n\tau_o}
  \nonumber\\
&&  \times(\sin\theta_0 \sin n\sigma-\cos\theta_0 \cos n\sigma) \quad \text{at} \quad \tau=\tau_o.\quad
  \label{eq7E11outerDirichlet}
\end{eqnarray}

The surrounding electric field exerts electrostatic force and/or moment on the cylinder,
\begin{equation}
\mathbf{F}=\int_A{\mathbf{\Pi}\cdot \mathbf{e}_\tau dA}, 
  \label{eq15-DEPforce}
\end{equation}
\begin{equation}
\mathbf{M}=R_i\int_A{\mathbf{e}_\tau\times(\mathbf{\Pi}\cdot\mathbf{e}_\tau)dA}.
  \label{eq16-DEPmoment}
\end{equation}

The electric field is expelled by the EDLs on the cylinder. Thus, only the tangential electric field $E_\sigma$ remains. The Maxwell stress tensor $\mathbf{\Pi}_e=\varepsilon_w(\mathbf{E}\mathbf{E}-\frac{1}{2}E^2\mathbf{I})$ on the cylinder surface is normal, $\mathbf{\Pi}_e\cdot \mathbf{e}_\tau=-\varepsilon_wE_\sigma^2\mathbf{e}_\tau/2$. Therefore, it can be concluded that the electrostatic moment $M_e$ is zero. Substituting Eq.\ (\ref{eq5phiDirichlet}) into Eq.\ (\ref{eq15-DEPforce}) through the Maxwell stress tensor, the electrostatic forces per unit length on the cylinder are obtained,
\begin{equation}
F_{e,x}=\pi\varepsilon_w E_0^2 a \sum_{n=1}^{\infty} \frac{2 n^2 \text{e}^{-2n\tau_o}}{\cosh^2 n(\tau_i-\tau_o)},
  \label{eq15FexD}
\end{equation}
\begin{equation}
F_{e,y}=0.
  \label{eq15FeyD}
\end{equation}

The 2D fluid flow is described by the biharmonic equation of the stream function $\psi$,
\begin{equation}
\nabla^4\psi=0
  \label{eq8-biharmonic},
\end{equation}
where $\psi$ is related to the velocities in the bipolar coordinates through,
\begin{equation}
u_\sigma=h\frac{\partial \psi }{\partial \tau },\quad u_\tau=-h\frac{\partial \psi }{\partial \sigma },
  \label{eq9-psiu}
\end{equation}
where $h=(\cosh\tau-\cos\sigma)/a$.

The general solution of the stream function $\psi$ was given by Jeffery \cite{jeffery1922rotation},
\begin{eqnarray}
  h\psi &=& A_0\cosh\tau+B_0\tau(\cosh\tau-\cos\sigma)+C_0\sinh\tau
  \nonumber\\
&& +D_0\tau\sinh\tau+\sum_{n=1}^{\infty}[a_n\cosh(n+1)\tau
\nonumber\\
&& +b_n\cosh(n-1)\tau+c_n\sinh(n+1)\tau
   \nonumber\\
&& +d_n\sinh(n-1)\tau]\cos n\sigma+h_1\tau\sin\sigma
\nonumber\\
&&+\sum_{n=1}^{\infty}[e_n\cosh(n+1)\tau+f_n\cosh(n-1)\tau
\nonumber\\
&&  +g_n\sinh(n+1)\tau+h_n\sinh(n-1)\tau]\sin n\sigma.\quad
  \label{eq14-psi}
\end{eqnarray}

For practical electrolyte concentrations ($10^{-6}\sim10^{-3}$ mol/L), the EDL thickness $\lambda_D$ ranges from nanometers to sub-micrometers. It is typically much smaller than the characteristic length of either natural colloidal systems or artificial microfluidic devices. Therefore, the thin EDL approximation is adopted ($\lambda_D\ll \text{min}(R_i, \, R_o-R_i)$). Under such condition, the electric field is coupled to the flow field through the Helmholtz-Smoluchowski formula,
\begin{equation}
\mathbf{u}_\sigma=-\frac{\varepsilon_w\zeta}{\mu}E_\sigma\mathbf{e}_\sigma,
 \label{eq0slipu}
\end{equation}
where $\varepsilon_w$ and $\mu$ are the dielectric permittivity and the viscosity of the electrolyte solution, respectively; $\zeta$ is the zeta potential, given as $\zeta_i$ (Eq.\ (\ref{eq7zetaiDirichlet})) and $\zeta_f$ for the conducting cylinder and the non-conducting cylindrical pore, respectively. Eq.\ (\ref{eq0slipu}) holds for hydrophilic surfaces. The systematic analysis on electrokinetic phenomena occurred on hydrophilic surfaces can be referred from Refs.\ \cite{yariv2009asymptotic,schnitzer2012macroscale}. For hydrophobic surfaces, the relationship between the slip velocity and the zeta potential alters. The detailed derivations can be referred from Ref.\ \cite{maduar2015electrohydrodynamics}.

Substituting Eqs.\ (\ref{eq7zetaiDirichlet}) and (\ref{eq7E11innerDirichlet}) into Eq.\ (\ref{eq0slipu}), the slip velocity on the cylinder is obtained. After mathematical manipulations, the boundary condition of fluid flow on the cylinder is expressed as,
\begin{eqnarray}
\mathbf{u}&=&\sum_{n=1}^{\infty}\left(K_{i,n}\cos n\sigma+\Lambda_{i,n}\sin n\sigma\right)\mathbf{e}_\sigma
  \nonumber\\
&& +U_x \mathbf{e}_x+U_y \mathbf{e}_y+\Omega R_i \mathbf{e}_\sigma \quad \text{at} \quad \tau=\tau_i,
  \label{eq16innerslipD}
\end{eqnarray}
where $K_{i,n}$ and $\Lambda_{i,n}$ are the coefficients of $\cos n\sigma$ and $\sin n\sigma$, respectively. Their expressions can be referred from Eqs.\ (S.12) $\sim$ (S.14) and (S.23) $\sim$ (S.24) of the Supplementary.

Substituting Eq.\ (\ref{eq7E11outerDirichlet}) and $\zeta_f$ into Eq.\ (\ref{eq0slipu}), the boundary condition of fluid flow on the cylindrical pore is obtained,
\begin{eqnarray}
\mathbf{u} &=&U_i \tilde{\zeta} \left[-\text{e}^{-\tau_o}\cos\theta_0+\sinh\tau_o\sum_{n=1}^{\infty}2\text{e}^{-n\tau_o}
  \right.\nonumber\\
&&\left. \times\left(\cos\theta_0\cos n\sigma-\sin\theta_0\sin n\sigma\right)\right]\mathbf{e}_\sigma \quad \text{at} \quad \tau=\tau_o.\quad\quad
  \label{eq15outerslipD}
\end{eqnarray}
where $U_i=\frac{\varepsilon_w E_0^2 R_i}{\mu}$ and $\tilde{\zeta}=\frac{\zeta_f}{E_0 R_i}$ are the velocity scale of the cylinder and the dimensionless zeta potential of the cylindrical pore, respectively.

The fluid flow can be decomposed into two parts according to its linearity. First, we consider the flow due to the electrokinetic slip velocities on the stationary cylinder and the cylindrical pore. The stream function $\psi$ of this part is determined by substituting Eq.\ (\ref{eq14-psi}) into the first term of Eq.\ (\ref{eq16innerslipD}) and Eq.\ (\ref{eq15outerslipD}) through Eq.\ (\ref{eq9-psiu}). The expressions of the coefficients are listed in Section A of the Supplementary. Substituting the stream function $\psi$ with the obtained coefficients into Eqs.\ (\ref{eq15-DEPforce}) and (\ref{eq16-DEPmoment}) through the viscous stress tensor $\mathbf{\Pi}_H=-p\mathbf{I}+\mu[\nabla\mathbf{u}+(\nabla\mathbf{u})^T]$, the hydrodynamic forces and moment per unit length on the cylinder are obtained as shown by Eqs.\ (S.25) $\sim$ (S.27) in Section A of the Supplementary. Next, we consider the flow due to the cylinder motion $U_x \mathbf{e}_x+U_y \mathbf{e}_y+\Omega R_i \mathbf{e}_\sigma$, which arouses drag forces and moment on the cylinder. The derivation of the drag forces and moment is presented in Section B of the Supplementary.

Since the cylinder is in free suspension, the net force and moment exerted on it should vanish. Sum up the obtained electrostatic, hydrodynamic, and drag forces along the $x-$axis, i.e., Eqs.\ (\ref{eq15FexD}), (S.25), (S.44); along the $y-$axis, i.e., Eqs.\ (\ref{eq15FeyD}), (S.26), (S.45); and the moments, i.e., Eqs.\ (S.27), (S.46), the cylinder velocities are obtained,
\begin{equation}
U_x =U_{ES,x}+U_{ICEO,x}+U_{EO,x},
\label{eq30UxD}
\end{equation}
where
\begin{equation}
\begin{split}
U_{ES,x} &=\frac{1}{2} U_i \sum_{n=1}^{\infty}\frac{n^2 \text{e}^{-2n\tau_o}\sinh\tau_i \left[\tau_i-\tau_o-\tanh(\tau_i-\tau_o)\right]}{\cosh^2 n(\tau_i-\tau_o)},
\end{split}\tag{22a}
\end{equation}
\begin{equation}
\begin{split}
U_{ICEO,x} &=\frac{1}{2} U_i \left\{\frac{\text{e}^{-2\tau_o} \sinh\tau_i \tanh(\tau_i-\tau_o)}{\cosh(\tau_i-\tau_o)}\left[\frac{\cos2\theta_0}{\cosh(\tau_i-\tau_o)}\right.\right.\\
&\left.\left.-\frac{2\text{e}^{-\tau_o}\cosh\tau_i}{\cosh2(\tau_i-\tau_o)}+\frac{\text{e}^{-2\tau_o}}{\cosh3(\tau_i-\tau_o)}\right]\right.\\
  & \left. +\sum_{n=2}^{\infty} \left[\frac{n\text{e}^{-2n\tau_o} \sinh\tau_i \tanh(\tau_i-\tau_o)}{\cosh n(\tau_i-\tau_o)}\right.\right.\\
&\left.\left. \times\left(\frac{2\text{e}^{\tau_o}\cosh\tau_i}{\cosh(n-1)(\tau_i-\tau_o)} -\frac{2\text{e}^{-\tau_o}\cosh\tau_i}{\cosh(n+1)(\tau_i-\tau_o)}\right.\right.\right.\\
  & \left.\left.\left. +\frac{\text{e}^{-2\tau_o}}{\cosh(n+2)(\tau_i-\tau_o)}\right)\right.\right.\\
&\left.\left.-\frac{(n+1)\text{e}^{-2n\tau_o} \sinh\tau_i \tanh(\tau_i-\tau_o)}{\cosh(n-1)(\tau_i-\tau_o)\cosh(n+1)(\tau_i-\tau_o)}\right]\right\},
\end{split}\tag{22b}
\end{equation}
\begin{equation}
\begin{split}
U_{EO,x} &=-U_i \tilde{\zeta} \text{e}^{-\tau_o}\tanh(\tau_i-\tau_o)\sinh\tau_o \sin\theta_0,
\end{split}\tag{22c}
\end{equation}
\begin{eqnarray}
U_y &=& -\frac{1}{2}U_i \frac{\text{e}^{-2\tau_o} \cosh\tau_i \tanh(\tau_i-\tau_o) \sin2\theta_0}{\cosh^2(\tau_i-\tau_o)}
\nonumber\\
&& +U_i \tilde{\zeta} \frac{\text{e}^{-\tau_o}\sinh(\tau_i-\tau_o)\left[1-\frac{\cosh\tau_i\sinh\tau_o}{\cosh(\tau_i-\tau_o)}\right]\cos\theta_0}{\sinh\tau_i},\nonumber\\
\label{eq30UyD}
\end{eqnarray}
\begin{eqnarray}
\Omega &=& -\frac{U_i}{R_i}\frac{\text{e}^{-2\tau_o}\tanh(\tau_i-\tau_o) \sin2\theta_0}{2\cosh^2(\tau_i-\tau_o)}
\nonumber\\
&& -\frac{U_i}{R_i} \frac{\tilde{\zeta}\text{e}^{-\tau_o}\sinh\tau_o\big[1+\tanh(\tau_i-\tau_o)\big]\cos\theta_0}{\sinh\tau_i}.
\label{eq30UrotD}
\end{eqnarray}

Clearly, the cylinder velocities are trigonometric functions of the electric field phase angle $\theta_0$. The cylinder velocities can be manipulated by changing $\theta_0$. Three factors contribute to the cylinder motion, namely the electrostatic (ES) force, the induced charge electroosmotic (ICEO) flow, and the electroosmotic (EO) flow. All these three factors contribute to $U_x$ (Eqs.\ (\ref{eq30UxD}) and (S.70)). Only the ICEO and the EO flows contribute to $U_y$ and $\Omega$ (Eqs.\ (\ref{eq30UyD}), (\ref{eq30UrotD}), (S.71) and (S.72)).

\section{Results and discussion}
\subsection{Cylinder velocities}\label{Secvelocities}
\begin{figure*}[!htb]
\centering
\includegraphics[width=6.5in]{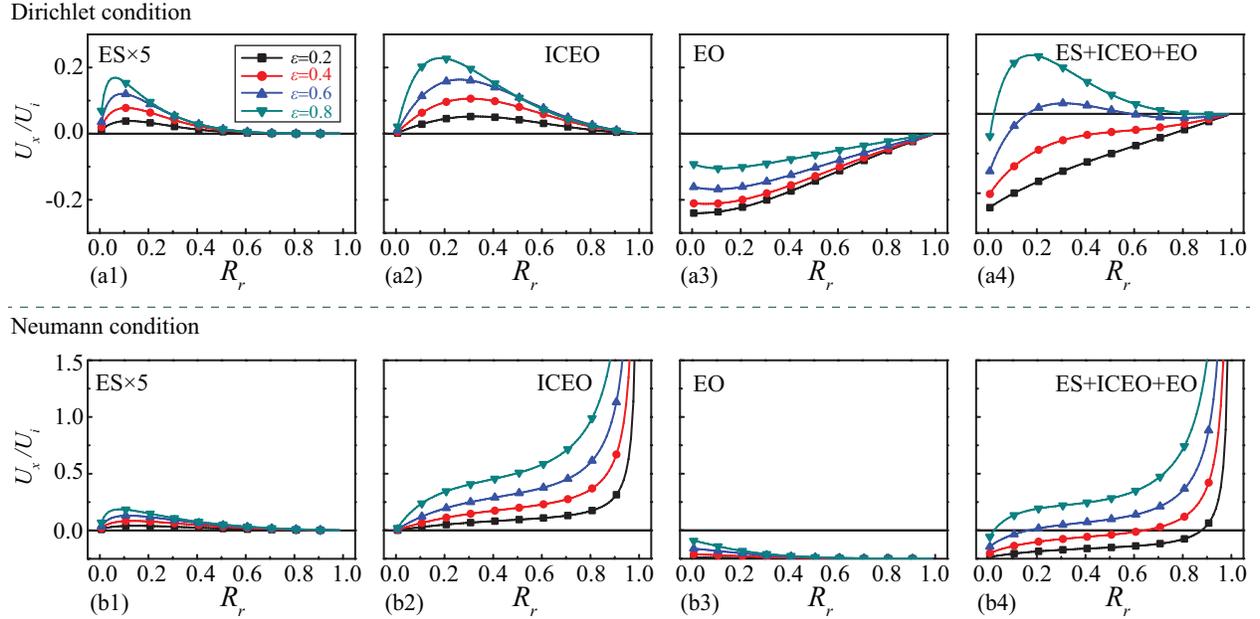}
\caption{Variation of the cylinder velocity $U_x/U_i$ with the radius ratio $R_r$ at different eccentricities $\varepsilon$. The ES component is 5 times amplified for a better observation.}
 \label{Fig2}
\end{figure*}

As introduced previously, three factors contribute to the cylinder velocities. The cylinder velocities due to these factors as well as the total velocities are characterized in this section. The influences of $R_r$ and $\varepsilon$ on the cylinder velocities are presented in Figs.\ \ref{Fig2} $\sim$ \ref{Fig4} with $\theta_0=\pi/6$ and $\tilde{\zeta}=1$. The influences of $\theta_0$ and $\tilde{\zeta}$ on the cylinder velocities are shown in Figs.\ S.1 $\sim$ S.3 of the Supplementary with $R_r=\varepsilon=0.5$. The ES component of cylinder velocity is irrelevant to $\tilde{\zeta}$ and $\theta_0$ (Fig.\ S.1). It solely contributes to $U_x$. As the induced zeta potential $\zeta_i$ is a function of the applied electric field, the ICEO component is irrelevant to $\tilde{\zeta}$ but is a trigonometric function of 2$\theta_0$ (Figs.\ S.1 $\sim$ S.3). The variations of cylinder velocities with $\tilde{\zeta}$ and $\theta_0$ follow the same trend in the two conditions (Figs.\ S.1 $\sim$ S.3).

\begin{figure*}[!htb]
\centering
\includegraphics[width=5in]{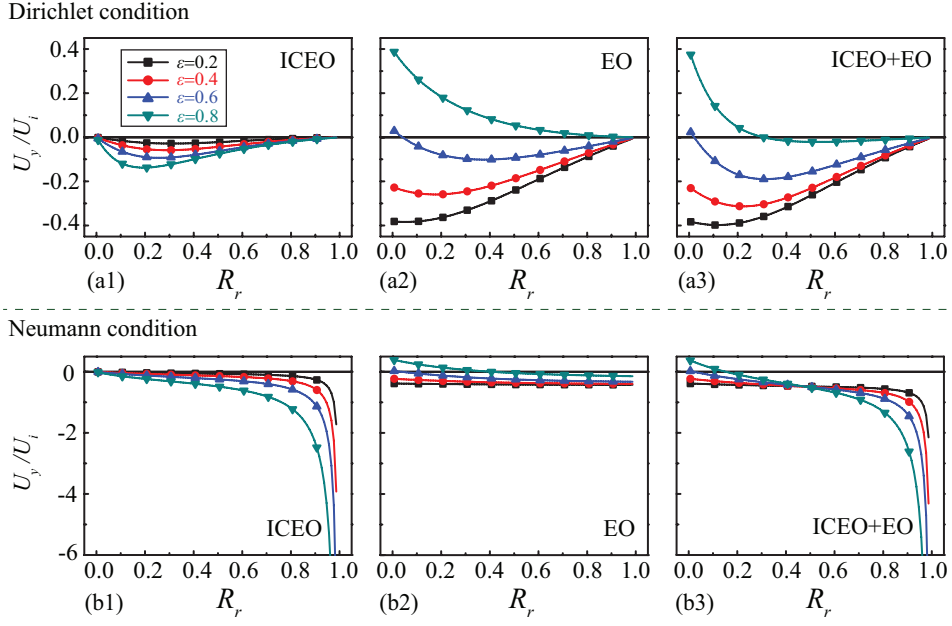}
\caption{Variation of the cylinder velocity $U_y/U_i$ with the radius ratio $R_r$ at different eccentricities $\varepsilon$.}
 \label{Fig3}
\end{figure*}

Fig.\ \ref{Fig2} shows the variation of the cylinder velocity $U_x$ with the radius ratio $R_r$ at different eccentricities $\varepsilon$. The ES components of $U_x$ obtained from the Dirichlet and the Neumann conditions follow the same trend as $R_r$ and $\varepsilon$ increase (Figs.\ \ref{Fig2}(a1) and \ref{Fig2}(b1)). They monotonically increase as $\varepsilon$ increases, and show a parabolic variation as $R_r$ increases. This is due to the fact that the ES component of $U_x$ is caused by the asymmetric surrounding electric field. A stronger asymmetry leads to a larger ES force. The asymmetry of the surrounding electric field monotonically increases as $\varepsilon$ increases. At $R_r=0$, the cylindrical pore is infinitely large compared to the cylinder. The cylindrical pore shows negligible influence on the local electric field around the cylinder. Thus, the ES force is zero. At $R_r=1$, the cylinder and the cylindrical pore coincide with each other. The surrounding electric field becomes totally symmetric, which leads to zero ES force. As $R_r$ increases from 0 to 1, the asymmetry of the surrounding electric field first increases and then decreases. The ES components of $U_x$ obtained from these two conditions are of the same order of magnitude although the Dirichlet condition leads to a faster decay as $R_r$ increases when $R_r$ is large. This trend can be more clearly observed in Fig.\ S.4(a).

Although the electric fields obtained from the two conditions do not show significant difference, the resulted cylinder velocities due to fluid flow show otherwise. Using the Dirichlet condition, the ICEO components of cylinder velocities ($U_x$, $U_y$ and $\Omega$) increase from zero and then diminishes to zero as $R_r$ increases (Figs.\ \ref{Fig2}(a2), \ref{Fig3}(a1) and \ref{Fig4}(a1)). While using the Neumann condition, they monotonically increase from zero as $R_r$ increases (Figs.\ \ref{Fig2}(b2), \ref{Fig3}(b1) and \ref{Fig4}(b1)). As $\varepsilon$ increases, the ICEO components of cylinder velocities obtained from both conditions show monotonic increases.

\begin{figure*}[!htb]
\centering
\includegraphics[width=5in]{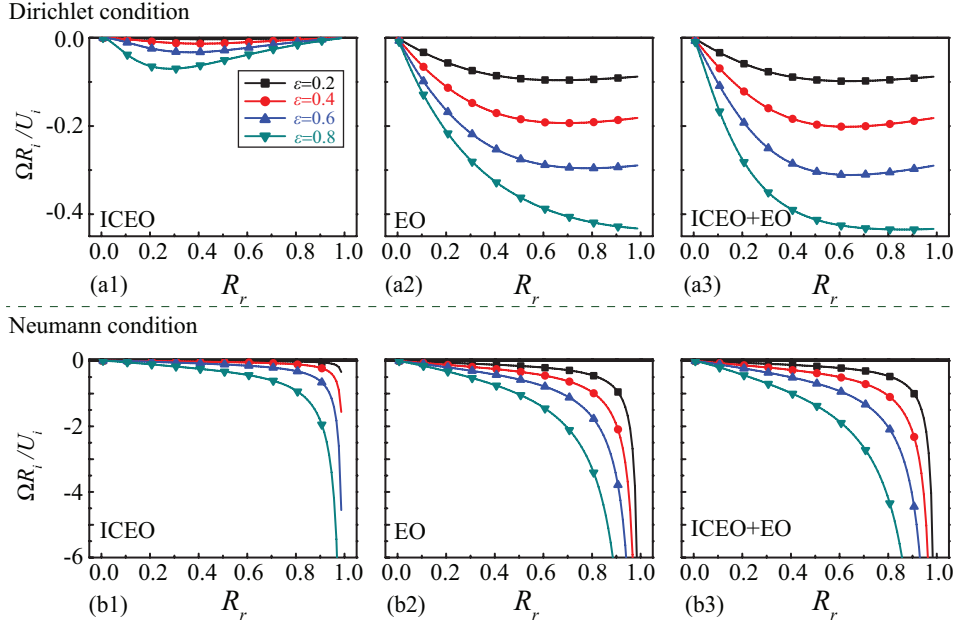}
\caption{Variation of the cylinder velocity $\Omega R_i/U_i$ with the radius ratio $R_r$ at different eccentricities $\varepsilon$.}
 \label{Fig4}
\end{figure*}

As $R_r$ increases, the EO components of $U_x$ and $U_y$ approach zero (Figs.\ \ref{Fig2}(a3) and \ref{Fig3}(a2)) and constant values (Figs.\ \ref{Fig2}(b3) and \ref{Fig3}(b2)) when they are obtained from the Dirichlet and the Neumann conditions, respectively. The EO components of $\Omega$ obtained from both conditions approach constant values as $R_r$ increases (Figs.\ \ref{Fig4}(a2) and \ref{Fig4}(b2)). In addition, as $\varepsilon$ increases, the magnitudes of the EO components of $U_x$ monotonically decrease (Figs.\ \ref{Fig2}(a3) and \ref{Fig2}(b3)), the EO components of $U_y$ increase from negative to positive (Figs.\ \ref{Fig3}(a2) and \ref{Fig3}(b2)), and the magnitudes of the EO components of $\Omega$ monotonically increase (Figs.\ \ref{Fig4}(a2) and (b2)).

From Figs.\ \ref{Fig2} $\sim$ \ref{Fig4}, we can conclude that the cylinder velocities obtained from the Neumann condition are larger than those obtained from the Dirichlet condition, especially at large $R_r$. One may refer to Figs.\ S.4 $\sim$ S.6 in Section D of the Supplementary for more detail. The Dirichlet and the Neumann conditions specify the electrical potential and the electric field (i.e., surface charge density) on the cylindrical pore, respectively \cite{jackson1999classical}. The tangential electric field on the cylindrical pore is not defined by the Neumann condition. Hence, the electric fields within the cylindrical pore vary due to the different boundary conditions. As $R_r$ reduces, the difference in these two conditions becomes more significant. The difference of the ICEO components obtained in the two conditions is more pronounced than that of the EO components. This is due to the fact that the ICEO component is a quadratic function of electric field, while the EO component is linearly proportional to electric field. Cylinder velocities depend on the relative magnitudes of their components as shown in Figs.\ \ref{Fig2} $\sim$ \ref{Fig4}.

\begin{figure}[!htb]
\centering
\includegraphics[width=6in]{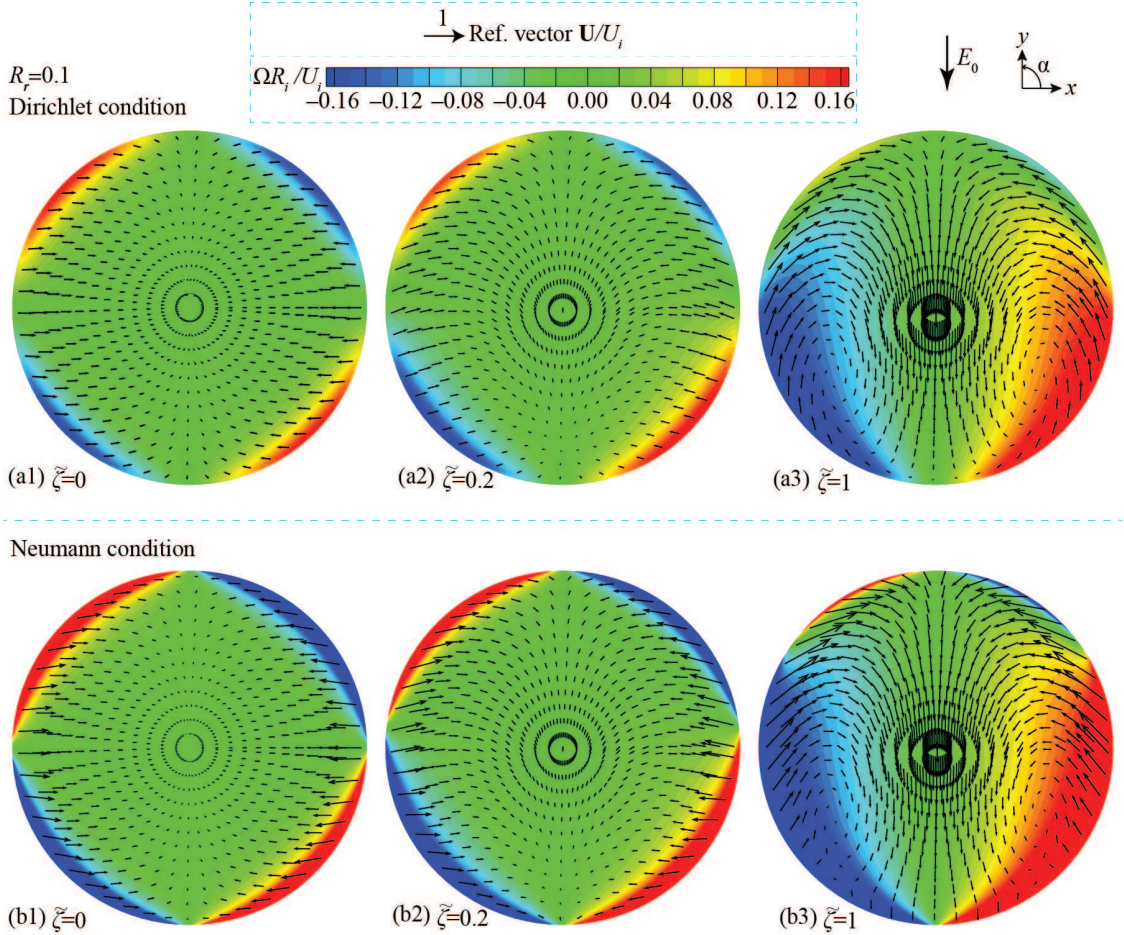}
\caption{Cylinder velocity maps. $\tilde{\zeta}=0$, 0.2 and 1 in Figs.\ (a1,b1), (a2,b2) and (a3,b3), respectively. Radius ratio $R_r=0.1$ and electric field phase angle $\theta_0=0$. The vector field indicates the cylinder translational velocity $\mathbf{U}=U_x\mathbf{e}_x+U_y\mathbf{e}_y$. The contour plot shows the cylinder rotational velocity $\Omega$, where the positive and negative values indicate the counterclockwise and clockwise directions, respectively.}
 \label{Fig5}
\end{figure}

The cylinder velocity maps at $\theta_0=0$ are shown in Fig.\ \ref{Fig5}. The vectors and contours indicate the translational and rotational velocities of the cylinder, respectively. For a nonzero $\theta_0$, the cylinder velocity map is the same as that at $\theta_0=0$ but tilts at an angle $\theta_0$. To facilitate the discussion, a polar coordinate system is introduced (as seen in the upper right corner of Fig.\ \ref{Fig5}), where $\alpha$ indicates the angle of the polar coordinates.

Both the translational and rotational velocities of the cylinder obtained from the Neumann condition (Fig.\ \ref{Fig5}(b)) are larger than those obtained from the Dirichlet condition (Fig.\ \ref{Fig5}(a)). The contour plot demonstrates that the cylinder possesses a greater rotational velocity $\Omega$ when it is near the cylindrical pore at $\alpha=(2n-1)\pi/4$. As $\tilde{\zeta}$ increases, the peak values of $\Omega$ at $\alpha=5\pi/4$ and $7\pi/4$ increase while those at $\alpha=\pi/4$ and $3\pi/4$ reduce or even disappear. This is due to the increasing EO component of $\Omega$. At a given electric field, the magnitude and the direction of $\Omega$ can be tuned by adjusting the position of the cylinder within the cylindrical pore.

The vector fields in Fig.\ \ref{Fig5} show the magnitude distributions of $\mathbf{U}$ and the cylinder trajectories. The cylinder experiences large $\mathbf{U}$ when it is close to the cylindrical pore. When $\tilde{\zeta}=0$, the results show that the cylinder moves towards and becomes stationary at the center of cylindrical pore regardless of its initial position (Figs.\ \ref{Fig5}(a1) and \ref{Fig5}(b1)). At $\tilde{\zeta}=0.2$, the cylinder moves towards and becomes stationary at a stationary point near the cylindrical pore due to the increased EO component of $\mathbf{U}$ (Figs.\ \ref{Fig5}(a2) and \ref{Fig5}(b2)). As $\tilde{\zeta}$ increases to 1, two stationary points appear within the cylindrical pore because the EO component of $\mathbf{U}$ is greatly enhanced (Figs.\ \ref{Fig5}(a3) and \ref{Fig5}(b3)). Regardless of the initial position and the specific trajectory, the cylinder moves towards and becomes stationary at the stationary points using the Dirichlet condition (Fig.\ \ref{Fig5}(a3)); while it may become stationary at the stationary points or reach the lower side of the cylindrical pore using the Neumann condition (Fig.\ \ref{Fig5}(b3)). These different cylinder trajectories are caused by the different EO components of $\mathbf{U}$ obtained from the two conditions.

\subsection{Micromotor}
Since the cylinder rotates when it is eccentric with respect to the cylindrical pore (Figs.\ \ref{Fig4} and S.3), micromotors can be developed by letting the cylinder free to rotate but not translate. The rotation of the cylinder may influence the establishment of EDL on the cylinder. To ensure the influence is negligible, the rotational velocity of the cylinder must be slow compared to the establishment of EDL. The characteristic time of the EDL formation is the charging time $t_c$, defined as $t_c=\kappa^{-1}R_i/D_i$. Here $\kappa^{-1}=\sqrt{\varepsilon_wk_B N_A T/(2F^2c_0)}$ is the Debye length of the EDL, where $\varepsilon_w$ is the dielectric permittivity of the electrolyte solution; $k_B$ is the Boltzmann constant; $N_A$ is the Avogadro constant; $T$ is the absolute temperature of the electrolyte solution; $F$ is the Faraday constant; and $c_0$ is the molar concentration of the electrolyte solution. The EDL charging time of the cylinder $t_c$ is larger than the Debye relaxation time for ionic screening, $t_D=\kappa^{-2}/D_i$, while smaller than the diffusion time for the relaxation of bulk concentration gradient, $t_R=R_i^2/D_i$, by a factor of $\kappa R_i$. $\kappa R_i$ is typically large in microfluidics \cite{squires2004induced}. We hereby define the charging frequency $f_c=1/t_c=\Gamma_c \frac{\sqrt{c_0}}{R_i}$, where $\Gamma_c=\sqrt{2}D_i F/\sqrt{\varepsilon_w k_B N_A T}$ is a constant with the given parameters in Table \ref{table1}. The charging frequency $f_c$ is proportional to $\sqrt{c_0}$ and inversely proportional to $R_i$. To ensure the effect of the cylinder rotation on the establishment of EDL is insignificant, the rotational velocity $\Omega$ of cylinder should be much smaller than the charging frequency $f_c$, $\Omega\ll f_c$.

\begin{table}[htb] 
\centering
\footnotesize
\caption{Parameters of the electrokinetic system of the electrolyte solution. \label{table1}}
\begin{tabular}{l l l}
\hline
Dielectric permittivity & $\varepsilon_w$ & $7\times 10^{-10}\;{\rm kg\cdot m\cdot V^{-2}\cdot s^{-2}}$\\
Viscosity               & $\mu$           & ${\rm 1\times 10^{-3}\; kg\cdot m^{-1}\cdot s^{-1}}$\\
Density                 & $\rho$          & ${\rm 1\times 10^{3}\; kg\cdot m^{-3}}$\\
Diffusivity             & $D_i$           & ${\rm 1\times 10^{-9}\; m^2\cdot s^{-1}}$\\
Boltzmann constant      & $k_B$           & $1.38\times 10^{-23}\; {\rm J\cdot K^{-1}}$ \\
Avogadro constant       & $N_A$           & $6.02\times 10^{23}\; {\rm mol^{-1}}$\\
Absolute temperature    & $T$             & $298.15\; {\rm K}$\\
Faraday constant        & $F$             & $9.65\times 10^{4}\; {\rm C\cdot mol^{-1}}$\\
\hline
\end{tabular}
\end{table}
The rotational velocity scale $\Omega_i=U_i/R_i=\Gamma_r E_0^2$ is used to represent cylinder rotation in the following analysis, where $\Gamma_r=\varepsilon_w/\mu$ is a constant with the given parameters in Table \ref{table1}. The present study is carried out with the quasi-steady state assumption, i.e., the unsteady term, $\rho\partial\mathbf{u}/\partial t$, in the Stokes equation is neglected. To ensure the validity of this assumption, the diffusion time of fluid vorticity, $t_\nu=R_i^2/\nu$, should be much smaller than the advection time scale of the flow, $t_a=R_i/U_i$ \cite{pozrikidis2011introduction}, where $\nu=\mu/\rho$ is the kinematic viscosity. Clearly, $\Omega_i=t_d^{-1}$, thus, $\Omega_i\ll t_\nu^{-1}$.

Given $c_0=1\times 10^{-3}$ mol$\cdot$L$^{-1}$, then $\kappa^{-1}\approx10$ nm. We take the cylinder radius $R_i=10\;\mu$m, thus, $\kappa R_i\approx1\times10^3$, the thin EDL approximation holds. And $t_c$ is much larger than $t_D$, while much smaller than $t_R$. The charging frequency of the EDL establishment $f_c\approx1\times10^4$ s$^{-1}$, which is much larger than the rotational velocity of most micromotors. And $\tau_\nu^{-1}=1\times10^4$ s$^{-1}$, the unsteady term in the Stokes equation is ensured negligible so long as $\Omega_i$ is much smaller than this value.

The rotational velocity of the load-free micromotor is the same as that shown in Figs.\ \ref{Fig4} and S.3. When the micromotor works under a load $M_l$, the moment balance becomes $M_e+M_H+M_d=M_l$. Substituting Eqs.\ (S.27) and (S.46) into this equation, the relationship between the rotational velocity of the micromotor $\Omega_m$ and the load $M_l$ is obtained,
\begin{eqnarray}
\Omega_m &=& \left\{\frac{2\tilde{\zeta}U_i}{R_i}\text{e}^{-\tau_o}\sinh\tau_o\sinh\tau_i\left[1-\left(\tau_i-\tau_o\right)\left(1+\coth(\tau_i-\tau_o)\right)
\right.\right.\nonumber\\
&&\left.\left.-\cosh\tau_o\sinh\tau_o-\coth\tau_i\sinh^2\tau_o\right]\cos\theta_0
\right.\nonumber\\
&&\left.+\frac{U_i}{2R_i} \frac{\text{e}^{-2\tau_o}\sinh\tau_i}{\cosh^2(\tau_i-\tau_o)}\left[\cosh(\tau_i-2\tau_o)-\cosh\tau_i+2(\tau_i-\tau_o)\sinh\tau_i\right]\sin2\theta_0
\right.\nonumber\\
&&\left.+\frac{M_l}{4\pi\mu R_i^2}\left[(\tau_i-\tau_o)(\cosh2\tau_i+\cosh2\tau_o-2)-\sinh2\tau_i+\sinh2\tau_o+\sinh2(\tau_i+\tau_o)\right]\right\}
\nonumber\\
&& /\left[\cosh2\tau_i-\cosh2(\tau_i-\tau_o)-2(\tau_i-\tau_o)\coth(\tau_i-\tau_o)\sinh^2\tau_i\right].
\label{eq31loadc}
\end{eqnarray}

\begin{figure*}[!htb]
\centering
\includegraphics[width=4.7in]{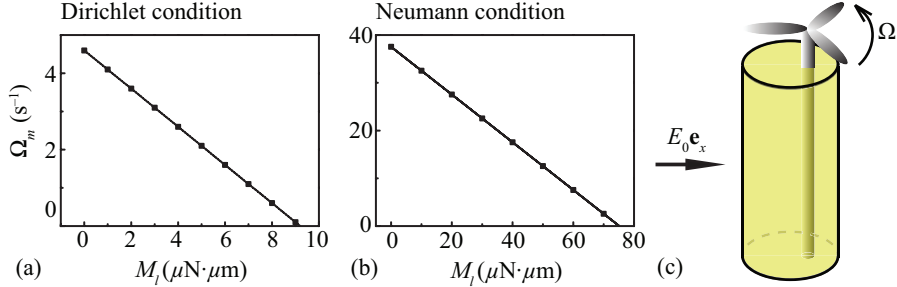}
\caption{Variation of the rotational velocity of the micromotor with the load $M_l$ obtained from (a) the Dirichlet condition and (b) the Neumann condition, and (c) schematic diagram of the micromotor. Here the radius ratio $R_r=0.5$, the eccentricity $\varepsilon=0.5$, the electric field phase angle $\theta_0=7\pi/4$, the dimensionless zeta potential of the cylindrical pore $\tilde{\zeta}=0$, the cylinder radius $R_i=10\,\mu$m, and the electric field strength $E_0=10$ kV$\cdot$m$^{-1}$.}
 \label{FIG10}
\end{figure*}

A schematic diagram of the micromotor and the variation of the rotational velocity of the micromotor $\Omega_m$ with the load $M_l$ are presented in Fig.\ \ref{FIG10}. $\Omega_m$ reduces linearly as $M_l$ increases and reaches zero when $M_l$ equals to the hydrodynamic moment $M_H$. The ranges of $M_l$ and $\Omega_m$ obtained from the Neumann condition is much larger than that obtained from the Dirichlet condition. By choosing the appropriate parameters according to Figs.\ \ref{Fig4} and S.3, a micromotor can be developed with a controllable rotational velocity. $M_H$ increases as the radius ratio $R_r$ and the eccentricity $\varepsilon$ increase (Eqs.\ (S.27) and (S.69)). Thus, the upper limit of the load $M_l$ can be increased by increasing $R_r$ and $\varepsilon$. Accordingly, a micromotor with a much faster rotational velocity $\Omega_m$ and a larger bearing capacity, i.e., a larger load $M_l$, can be developed. The micromotor can also bear loads in the opposite direction by adjusting the electric field phase angle $\theta_0$.

Both the Dirichlet and the Neumann boundary conditions of the electric field have been used in the studies of particle suspensions \cite{levine1974prediction,kozak1989electrokinetics,ohshima1997electrophoretic}. The different results are due to the fact that the Dirichlet boundary condition (Eq.\ (\ref{eq4potentialDirichlet})) defines the electrical potential on the cylindrical pore, while the Neumann boundary condition (Eq.\ (\ref{eq4potentialNeumann})) defines the surface charge density \cite{jackson1999classical}. Eq.\ (\ref{eq4potentialNeumann}) does not specify the tangential electric field on the cylindrical pore. It was reported that the statistical mechanics modelling on the electrophoresis of biocells favors the Dirichlet boundary condition \cite{keh2000diffusiophoresis}.

\section{Conclusion}
In this paper, the induced charge electrophoresis of a conducting cylinder suspended in a non-conducting cylindrical pore is theoretically studied, and a micromotor is proposed utilizing the cylinder rotation. Both the Dirichlet and the Neumann boundary conditions of the electric field are applied on the cylindrical pore. The analytical study on the cylinder velocities shows that the cylinder not only translates but also rotates when the cylinder and the cylindrical pore are eccentric. The cylinder velocities are examined with various values of the eccentricity, the radius ratio, the electric field phase angle, and the zeta potential of the cylindrical pore. The analysis shows that the cylinder velocities obtained in the two boundary conditions present great differences. Moreover, the cylinder trajectories show that the cylinder always approaches and becomes stationary at certain stationary points within the cylindrical pore.

By choosing the appropriate parameters of the electrokinetic system, the micromotor proposed in this paper can achieve a high rotational velocity without influencing the EDL establishment on the cylinder. A large eccentricity and a strong electric field are preferred to develop a micromotor with a high rotational velocity and a great bearing capacity. The micromotor proposed here has advantages of simple geometry and low operating voltage, and a great potential in the application of the lab-on-a-chip systems for chemical and biological analysis.

\section*{Acknowledgement}
The authors gratefully acknowledge research support from the Singapore Ministry of Education Academic Research Fund Tier 2 research Grant MOE2011-T2-1-036.

\bibliography{motorRef}

\end{document}